\documentclass[nouppercase]{ifmbe}
\usepackage{amsmath,esint}
\title{Ultra-sensitive SQUID instrumentation for MEG and NCI by ULF MRI}

\affiliation{Physikalisch-Technische Bundesanstalt, 10587 Berlin, Germany }{FIRSTAFF}

\author{R. K\"orber}{FIRSTAFF}

\begin{document}

\maketitle

\begin{abstract}
The requirements for the construction of ultra-sensitive SQUID instrumentation as used in biomagnetism are presented. Typically, SQUIDs are inductively coupled to pick-up coils and for this arrangement one can improve the noise performance by increasing the sensing coil area. To achieve optimum sensitivity one has to consider the signal-to-noise ratio (SNR), which is an intricate interplay between source characteristics and noise origin. It turns out that separate pick-up coil designs are needed for various noise characteristics even for an identical source. Hence, a hybrid system with differently sized pick-up coils presents the best option for multipurpose applications. A single channel system with close to SQUID intrinsic noise level is also described. This is possible by utilizing a special dewar design and thereby enabling a further increase in SNR. Such a system might be used for current density imaging and neuronal current imaging by ultra-low-field magnetic resonance where it also must be able to be field-tolerant to up to 100-200 mT.

\end{abstract}


\section{Introduction}
The use of superconducting quantum interference devices (SQUIDs) in biomagnetism is well established and commercial systems are available to enable the study of the magnetic field of the human brain in magnetoencephalography (MEG), for instance. Those devices are commonly based on low critical temperature (low-$T_{c}$) SQUIDs and have hundreds of channels. Their sensitivity is usually limited by thermal noise in the superinsulation of the liquid Helium (LHe) dewar to about 2~fT\, Hz$^{-1/2}$~\cite{Clarke2006}. 
\par Recently, the use of SQUIDs as NMR sensors has emerged in the field of ultra-low-field magnetic resonance (ULF MR)~\cite{ULFNMR2014}. Here, the SQUIDs are usually designed as current sensors and inductively coupled to a superconducting pick-up coil. In ULF MR, the sample is first magnetized in a strong polarizing field (up to 100-200~mT). It then relaxes or precesses in a much lower detection field of the order of $\mu$T. The use of the polarizing field places strong constraints on the design of the SQUID system as a whole necessitating the use of special precautions such as current limiters in the input circuit and the use of superconducting shields. Further, the currently relatively poor signal-to-noise ratio (SNR) in ULF MR results in a long averaging time and a voxel size in the mid mm range.
\par On the other hand, ULF MR offers the potential advantage of combining MEG and MRI into one instrument thereby improving the localization accuracy of MEG by minimizing co-registration errors. If combined with current density imaging (CDI) the conductivity map of the cortex could be extracted which would lead to further improvement. Other techniques presently investigated include neuronal current imaging (NCI) which aims at detecting directly the minute effects of neuronal currents in an MR image. 

\par This paper discusses possible steps to improve the sensitivity in SQUID detected MR. Using current sensors coupled to a pick-up coil, one can improve the field resolution by increasing the pick-up coil area. However, field resolution is not the only parameter of interest, instead it is the SNR which is of ultimate importance. As we will see below, this turns-out to be and intricate problem depending on the source and noise characteristics of the entire experimental setup necessitating either an optimized system for a particular problem or hybrid systems for multipurpose applications.

\section{Materials and methods}
\subsection{Signal-to-Noise ratio calculation}

In order to optimize the pick-up loop diameter for different applications, it is necessary to have a sound knowledge of the underlying source and noise characteristics. As shown below, there is no universal pick-up coil dimension for different applications. The following analysis is restricted to magnetic moments as they appear in ULF MR. 

\par Assume a magnetic moment \textbf{m} arising from a fully magnetized voxel of 1~mm$^3$ size, taking the spin density of water and a polarization field of 100~mT. The moment points along the vertical z-direction centrally underneath a circular pick-up coil of inductance $L_{p}$. Its signal flux $\Phi_{S}$ is calculated in dependence of the source depth $z$ and the coil diameter $d$. One can use the magnetic vector potential \textbf{A} of a magnetic moment \textbf{m}:

\begin{equation}
\textbf{A}(\textbf{r})=\frac{\mu_{0}}{4\pi}\frac{\textbf{m}\times\textbf{r}}{|\textbf{r}|^{3}}.
\end{equation} 

\par In this rotationally symmetric geometry there is only an azimuthal component $A_{\phi}$ of the magnetic vector potential \textbf{A} at the pick-up coil wire, hence $\Phi_{S}$, given by the closed line integral of \textbf{A}, is simply obtained by:
\begin{equation}
\Phi_{\textrm{S}}=\oint{\textbf{A}\textrm{d}\textbf{l}}=A_{\phi}d\pi.
\end{equation}

\par If the magnetic moment is not situated centrally under the pick-up coil, $\Phi_{S}$ is obtained via numerical integration. The conversion into the signal flux in the SQUID $\Phi_{SQ}$ assumes perfect matching for every pick-up coil diameter, i.e. the input coil inductance $L_{i}$ equals $L_{p}$, a SQUID inductance $L_{SQ}$ of 80~pH, and a coupling constant $k\approx 0.7$ applicable for the double transformer scheme implemented in the PTB SQUIDs~\cite{Drung2007}. This leads to:

\begin{equation}
\Phi_{SQ}=\Phi_{S}\frac{k\sqrt{L_{SQ}L_{i}}}{L_{\textrm{tot}}},
\end{equation}
where $L_{\textrm{tot}}$ is the total inductance of the input circuit.

\par In order to evaluate the noise contributions, a SQUID intrinsic flux noise of 0.6~$\mu\Phi_{0}$\,Hz$^{-1/2}$ is used. For a more realistic setup, dewar noise is also included. Specifically, 1.28~fT\,Hz$^{-1/2}$ as measured with a 17.1~mm diameter pick-up coil is chosen (see below). In this configuration, correlations within the pick-up coil leading to a diameter dependent flux noise are also taken into account~\cite{Nenonen1996}. The SNR is then calculated for the two cases.

\subsection{Hybrid multichannel system}

\begin{figure}[t]
      \centering
          \includegraphics[width=0.95\columnwidth]{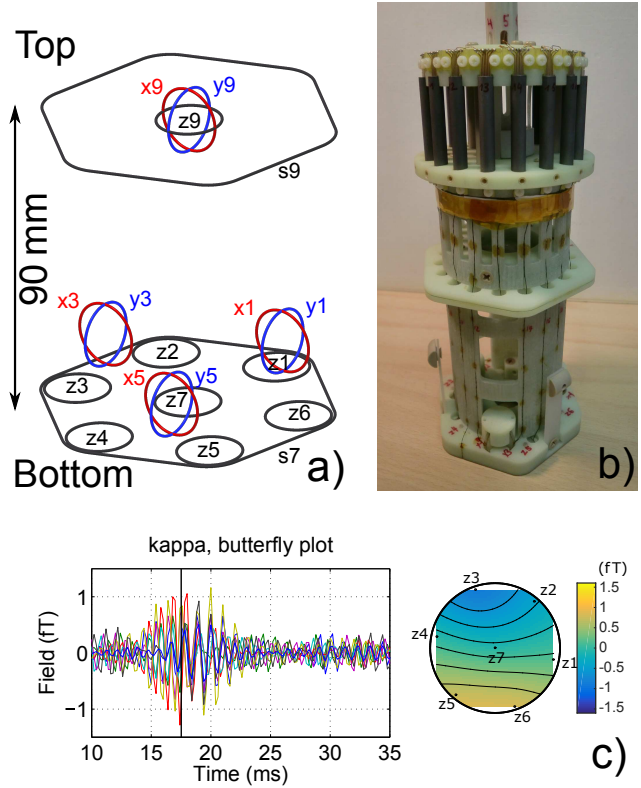}
      \caption{a) Schematic layout of the hybrid multichannel SQUID system. b) Picture of the module and c) Example of a 1~kHz burst (kappa) measurement evoked by electrostimulation of the median nerve. Note the minute amplitude of about 1 fT. Adapted from~\cite{Storm2016}.}
      \label{figure1}
\end{figure}

As we will see from the results of the SNR calculation, a hybrid multichannel system with various pick-up coil diameters allows one to achieve optimum sensitivity for different sources encountered in various applications such as MEG and/or ULF MR. Consequently, we built an 18 channel SQUID system equipped with two different pick-up coil diameters and orientations~\cite{Storm2016} as shown in Fig.\ref{figure1}. 

\par In the bottom plane and sensitive in the $z$-direction, a large hexagonal shaped pick-up coil with effective diameter 74.5~mm surrounds 7 smaller circular pick-up coils of diameter 17.1~mm with a center-to-center distance of 30~mm. Hence, conventional MEG for sources on the cortex is also possible as the small coils fulfill the Nyquist criteria to avoid undersampling in the spatial frequency domain~\cite{Ahonen1993a}. In addition, close to the bottom plane, there are also 3 $xy$-coil pairs of 17.1~mm diameter to detect transverse field components. The top level, 90~mm above the bottom plane, contains a large coil and a $xyz$-coil triplet which can be used for synthetic gradiometers. The system is currently operated inside a commercial low noise dewar with a warm-cold distance of 28~mm. The measured white field noise in the bottom plane amounts to 0.61 fT\,Hz$^{-1/2}$ and 1.28 fT\,Hz$^{-1/2}$ for the large and the small magnetometer, respectively, which formed the basis of the SNR simulations described above.

\section{Results}

\begin{figure}[t]
      \centering
          \includegraphics[width=0.95\columnwidth]{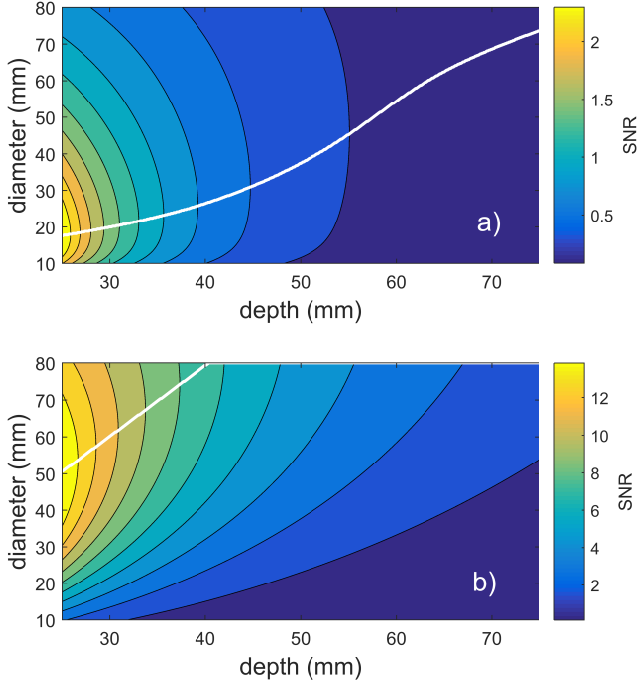}
      \caption{Signal-to-Noise ratio for a) SQUID and dewar noise and b) SQUID noise only. The magnetic moment is located centrally under the pick-up coil. The white line gives the optimum diameter for a given depth.}
      \label{figure2}
\end{figure}

\par First, the results of the SNR calculation for a magnetic moment located centrally underneath the pick-up coil are presented. Fig.\ref{figure2}~a) shows the realistic setup, i.e. including dewar noise, together with the optimum coil diameter in dependence of the source depth $z$ (white line). For $z=25$~mm an optimum diameter of about 17~mm with an SNR of 2.4 is obtained, for larger depth $z$ the optimum is very shallow resulting in $d\approx z$ for the optimum.

\par Fig.\ref{figure2}~b) gives the SNR if only SQUID noise is taken into account. Obviously, in this case a larger diameter is more sensitive for a given depth: For $z=25$~mm the optimum diameter is about 51~mm with an SNR of approximately 15. It is also worth noticing that the absolute SNR is about a factor of 10 larger for $z>30$~mm compared to the setup including dewar noise.

\par That the inclusion of dewar noise leads to a smaller optimum diameter can also intuitively be understood. For a given depth $z$ and a sufficiently small $d$, the SQUID noise is the main contribution initially. On increasing $d$, successively more external field noise is collected which ultimately dominates. 

\par It is also instructive to consider the influence of the dewar noise on the total SNR as shown in Table~\ref{table1}. Here, the total SNR is calculated for the large coil and the combination of the 7 small coils as applicable to our hybrid system. Two cases are considered in detail. First, signals from just outside the multichannel dewar with a representative depth $z$ of 30~mm. For signals originating from the cortex, a skull thickness of about 15-20~mm can be assumed giving a depth of about 50~mm.

\begin{table}[t]
  \footnotesize \onehalfspacing
  \caption{Total SNR for two setups and two representative source depths $z$. The diameter $d$ and the center-to-center distance $D$ of the small pick-up coils are applicable to the actual hybrid system.}
  \begin{tabular}{p{2.20cm}p{1.50cm}p{1.50cm}p{1.50cm}}
     \hline
     \normalfont Setup & depth (mm) & SNR SQUID noise & SNR SQUID $\&$ dewar noise\\
     \hline
     $1\times d=74.5$~mm & 30 & 11.2 &  0.76\\
     $7\times d=17.1$~mm, $D$=30~mm & 30 & 4.21 & 1.58 \\
    $1\times d=74.5$~mm & 50 & 5.04 &  0.34\\
     $7\times d=17.1$~mm, $D$=30~mm & 50 & 1.32 & 0.50 \\ 	
     \hline
  \end{tabular}
  \label{table1}
\end{table}

\par If only SQUID noise is present the large coil always outperforms the smaller coils by more than a factor of 2 improving with increasing source depth. If dewar noise is present, the situation is reversed and the combination of the 7 small coils performs better. Hence, a hybrid system offers the advantage of combining various coils giving flexibility for changing source and noise conditions. As an example, the low noise performance of our hybrid system enabled the detection of high frequency components, so-called $\sigma$ and $\kappa$-bursts, of somatosensory evoked activity by MEG as shown in Fig.\ref{figure1}~c). 

\section{Discussion}

\par The results of the calculation described above have a number of implications. First, even for a SQUID noise limited setup, there is no universal pick-up coil design enabling maximum SNR for all sources.

\par
\begin{figure}[t]
      \centering
          \includegraphics[width=0.95\columnwidth]{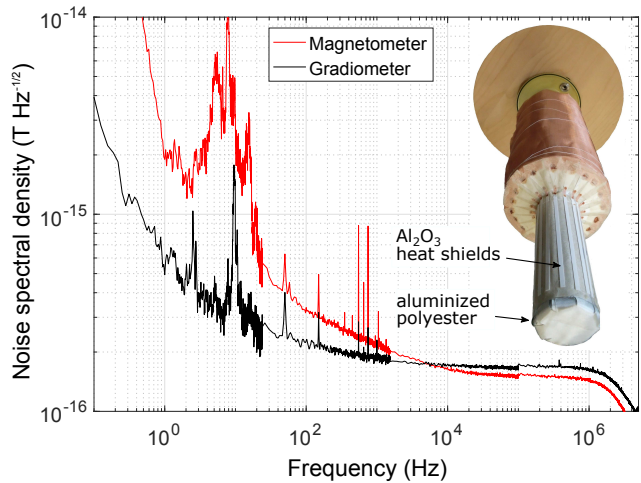}
      \caption{Measured magnetic flux density noise $S_{B}^{1/2}$ for the magnetometer and gradiometer pick-up coil with 45~mm diameter. Also shown is view of inside of the dewar highlighting the Al$_{2}$O$_{3}$ heat shields and the aluminized polyester. Adapted from~\cite{Storm2017}}
      \label{figure3}
\end{figure}

 The second implication is that a significant gain in SNR can be achieved if a dewar with negligible noise contributions could be used. This is particularly the case for larger pick-up coils. Hence, based on the design of Seton $et al.$\cite{Seton2005} we also built an ultra-low noise dewar. The use of aluminum oxide as heat shield material and aluminzed polyester as superinsulation reduces the dewar noise contribution to a negligible level. The dewar is currently equipped with a single channel SQUID system and has a warm cold distance of about 13~mm.  We achieve a white noise of about 150~aT\,Hz$^{-1/2}$ for a 45~mm diameter magnetometer pick-up coil~\cite{Storm2017}. For a 45~mm diameter first order axial gradiometer the measured noise amounts to about 170~aT\,Hz$^{-1/2}$ as shown in Fig.~\ref{figure3}. The noise is limited by the SQUID intrinsic flux noise with significant contributions from the read-out electronics. This setup is close to maximum SNR for a depth of 25~mm but is also suitable for the detection of deeper sources.

\par We intend to use this single channel system for the demonstration of current density imaging of impressed currents in the cortex. With the reduced warm-cold distance of $\sim$13~mm, the cortex is at a depth of about 30~mm below the pick-up coil and hence close to the optimum depth of our setup. The other application we are pursuing is neuronal current imaging (NCI) where the effect of long lived neuronal magnetic fields on the MR image would be detected. A suitable activity evoked by electrostimulation of the median nerve was identified with a depth of about 35~mm from the head surface~\cite{Koerber2013} giving an overall distance of $\sim$50~mm to the pick-up coil. Phantom studies demonstrating the feasibility of NCI showed a lack of SNR of about 3 for a less sensitive version of the system. Using the same 45~mm diameter gradiometer operating a noise level of about 1 fT\,Hz$^{-1/2}$ was achieved~\cite{Koerber2016}. Hence, with the improved setup NCI should in principle be possible.

\section{Conclusion}
In order to build a multipurpose SQUID system which can be used for diverse applications such as MEG and ULF MR it is best to use a hybrid design consisting of an array of smaller coils surrounded by large coils. The small coils can be used for MEG if designed to fulfill the Nyquist criteria for sampling the spatial frequency domain. For ULF MR the voxel depth and the noise characteristics determine the optimum pick-up coil diameter. A large coil will perform better in terms of SNR if one is limited by the intrinsic SQUID noise. This can be achieved as was shown very recently enabling a total measured field noise of around 170 aT\,Hz$^{-1/2}$ for a 45\,mm diameter first order gradiometer containing significant noise contribution from the electronics used for read out. This ultra-low noise performance should enable novel neuroimaging techniques such as neuronal current imaging.

\section*{Conflict of Interest}
The author declares that he has no conflict of interest.

\section*{Acknowledgements}

This project received funding from the \textit{European
Union's Horizon 2020 research and innovation programme}
under Grant Agreement No. 686865 and by the DFG under Grant No. KO 5321/1-1.

\bibliography{bibliography}

\end{document}